\documentclass[prl,aps,amssymb,twocolumn,groupedaddress,notitlepage,superscriptaddress]{revtex4-1}
\usepackage{amsmath}
\usepackage{amssymb}
\usepackage{amsthm}
\usepackage{amsfonts}
\usepackage{listings}
\usepackage{physics}
\usepackage[normalem]{ulem}
\usepackage{enumerate}
\usepackage{latexsym}
\usepackage{psfrag}
\usepackage{bm}
\usepackage{graphicx}
\usepackage[caption=false]{subfig}
\usepackage{blkarray}
\usepackage{array}
\usepackage{color}
\usepackage[normalem]{ulem}
\usepackage{multirow}

\usepackage[dvipsnames]{xcolor}
\definecolor{goodgreen}{rgb}{0.1,0.5,0}
\definecolor{goodred}{rgb}{0.7,0,0}
\usepackage[colorlinks,urlcolor=goodgreen,citecolor=blue,linkcolor=goodred]{hyperref}

\begin{document}
\title{On the robustness of topological corner modes in photonic crystals}
\author{Matthew Proctor}
\affiliation{%
Department of Mathematics, Imperial College London, London, SW7 2AZ, U.K.}
\author{Paloma Arroyo Huidobro}%
 \email{p.arroyo-huidobro@lx.it.pt}
\affiliation{%
Instituto de Telecomunica\c c\~oes, Instituto Superior Tecnico-University of Lisbon, Portugal}%

\author{Barry Bradlyn}%
\email{bbradlyn@illinois.edu}
\affiliation{Department of Physics and Institute for Condensed Matter Theory, University of Illinois at Urbana-Champaign, Urbana, IL, 61801-3080, USA}%

\author{Mar\'{i}a Blanco de Paz}
\affiliation{Donostia International Physics Center, 20018 Donostia-San Sebasti\'an, Spain}
\author{Maia G. Vergniory}
\affiliation{Donostia International Physics Center, 20018 Donostia-San Sebasti\'an, Spain}
\affiliation{IKERBASQUE, Basque Foundation for Science, Maria Diaz de Haro 3, 48013 Bilbao, Spain}
\author{Dario Bercioux}
\affiliation{Donostia International Physics Center, 20018 Donostia-San Sebasti\'an, Spain}
\affiliation{IKERBASQUE, Basque Foundation for Science, Maria Diaz de Haro 3, 48013 Bilbao, Spain}
\author{Aitzol Garc\'{i}a-Etxarri}
\email{aitzolgarcia@dipc.org}
\affiliation{Donostia International Physics Center, 20018 Donostia-San Sebasti\'an, Spain}
\affiliation{IKERBASQUE, Basque Foundation for Science, Maria Diaz de Haro 3, 48013 Bilbao, Spain}
\date{\today}






\begin{abstract}
We analyze the robustness of corner modes in topological photonic crystals, taking a $C_6$-symmetric breathing honeycomb photonic crystal as an example. First, we employ topological quantum chemistry and Wilson loop calculations to demonstrate that the topological properties of the bulk crystal stem from an obstructed atomic limit phase. 
We then characterize the topological corner modes emerging within the gapped edge modes employing a semi-analytical model, determining 
the appropriate real space topological invariants. For the first time, we provide a detailed account of the effect of long-range interactions on the topological modes in photonic crystals, and we quantify their robustness to perturbations. We conclude that, while photonic long-range interactions inevitably break chiral symmetry, the corner modes are protected by lattice symmetries. 

\end{abstract}

\maketitle


\textit{Introduction.|} Photonic topological insulators host protected boundary modes that are robust against a range of defects and imperfections~\cite{Ozawa2019}. While the paradigmatic case of two-dimensional (2D) topological photonic crystals (PhCs) hosting one-dimensional (1D) edge modes immune to back-scattering has been extensively studied~\cite{lu2014topological}, a hierarchy of protected boundary states of lower dimensionality are possible in higher-order topological insulators (HOTIs)~\cite{benalcazar2017quantized}. For instance, quantized quadrupole insulators in 2D, which were introduced in a generalization of the Su-Schrieffer-Heeger (SSH) model to a square lattice with a flux~\cite{benalcazar2017quantized}, host 1D edge states, as well as zero-dimensional (0D) corner modes. These higher-order topological modes (HOTMs) localized at the 0D corners of a 2D lattice benefit from topological protection. Just as HOTIs in condensed matter systems are characterized by charge fractionalization due to a filling anomaly of the bulk states~\cite{Kempkes2019,wieder2020strong,benalcazar2018quantization,zhu2020identifying}, classical wave HOTIs reveal an analogous fractional corner anomaly of the density of states~\cite{peterson2020fractional}. In systems with short-range hoppings and approximate chiral symmetry, these corner modes are mid-gap states~\cite{ssh1979, Asboth2016}.

HOTMs have been realised in a variety of classical systems including PhCs~\cite{ota2018photonic,xie2019visualization,chen2019direct,Li2020}, coupled photonic waveguides~\cite{Noh2018,Mittal2019,ElHassan2019}, 
phononic crystals~\cite{serra2018observation}, acoustic systems~\cite{Ni_2017,Ni2019,qi2020acoustic}, elastic systems~\cite{Fan2019elastic} and microwave circuits~\cite{peterson2018quantized}, and their robustness has been exploited for stable lasing~\cite{kim2020lasing, han2020lasing, gong2020topological}. 
However, 
a rigorous study of the effect of long-range interactions (the coupling between elements) which is unavoidable in many 
photonic systems \cite{koenderink2006complex,Pocock2018, Pocock2019, Li2020}, as well as a detailed analysis of the robustness of the HOTMs has not been undertaken. 
Here we consider a PhC with a $C_6$-symmetric lattice~\cite{Wu2015, Noh2018}, and fill the aforementioned gap by taking advantage of a semi-analytical model with long-range interactions~\cite{abajo2007colloquium}, that is, interactions beyond nearest neighbours between all the lattice elements. This allows us to perform an extensive study of the robustness of these modes against defects and imperfections. Crucially, we show that the HOTMs are protected by lattice symmetries; we quantify their degree of robustness against chiral-symmetry breaking long-range interactions, as well as to strong defects.

\begin{figure}
    \includegraphics[width=\columnwidth]{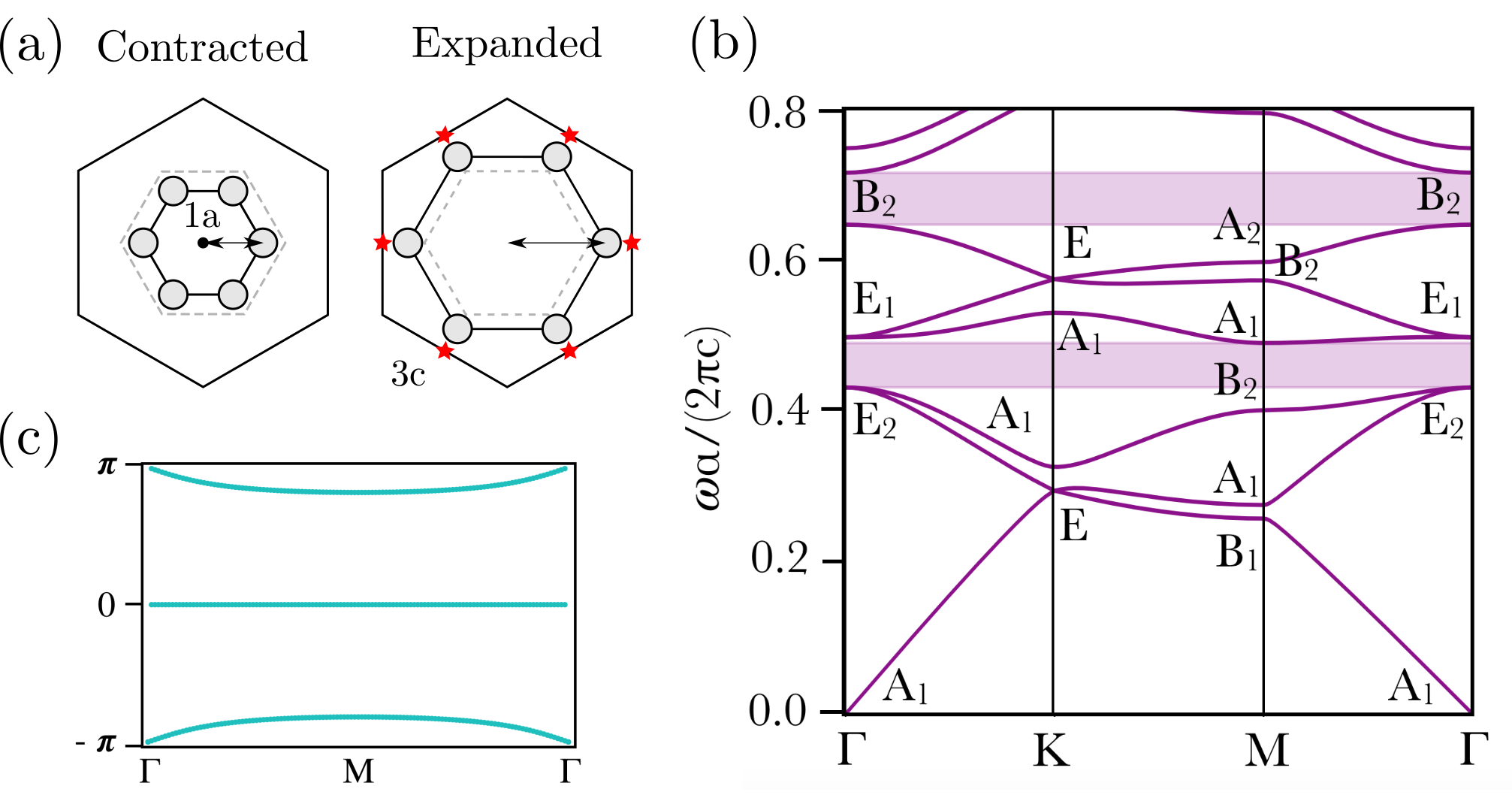}
    \caption{(a) Unit cells of the bulk lattice in the contracted and expanded phases, characterised by a contraction/expansion parameter, $\delta$. The relevant Wyckoff positions are labelled 1a (black circle) and 3c (red star). (b) Band structure of the silicon photonic crystal in the expanded phase for the TM modes. 
    The expansion coefficient $\delta$ is 0.11, and the radius of the cylinders is $0.12 a_0$, $a_0$ being the lattice constant of the crystal.
    (c) Wilson loops for bands 4 to 6 (Wilson loops for bands 1 to 3 are similar, see~\cite{supplemental}). }
    \label{fig:photonic_crystal_bulk}
\end{figure}


\textit{Photonic crystal.|}
We consider the breathing honeycomb PhC 
introduced in Ref.~\cite{Wu2015}, Fig.~\ref{fig:photonic_crystal_bulk}(a). Each unit cell in the triangular lattice consists of six silicon rods ($\varepsilon=11.7$) in vacuum of radius $r=0.12a_0$ located at a distance $R = R_0(1 \pm \delta)$ from the origin of the unit cell. Here, $a_0$ is the lattice parameter, and $R_0=a_0/3$ the location of the rods in the unperturbed honeycomb arrangement. The perturbation of the honeycomb lattice of rods by $\pm \delta$ yields expanded and contracted phases, respectively, where the doubly degenerate Dirac point at $\Gamma$ splits and a bulk band gap opens between $\omega a/(2\pi c) = 0.4$ to $0.5$. 
Although this band gap hosts 1D edge states as measured in several photonic experiments~\cite{barik2018topological,gorlach2018far,peng2019,Smirnova2019,parappurath2020topological,Liu2020photonic,Yang2020SpinMomentum}, we now discuss how they are not an instance of a $\mathbb{Z}_2$ topological insulator~\cite{Fragile2017}. 

\begin{figure*}
    \includegraphics[width=\textwidth]{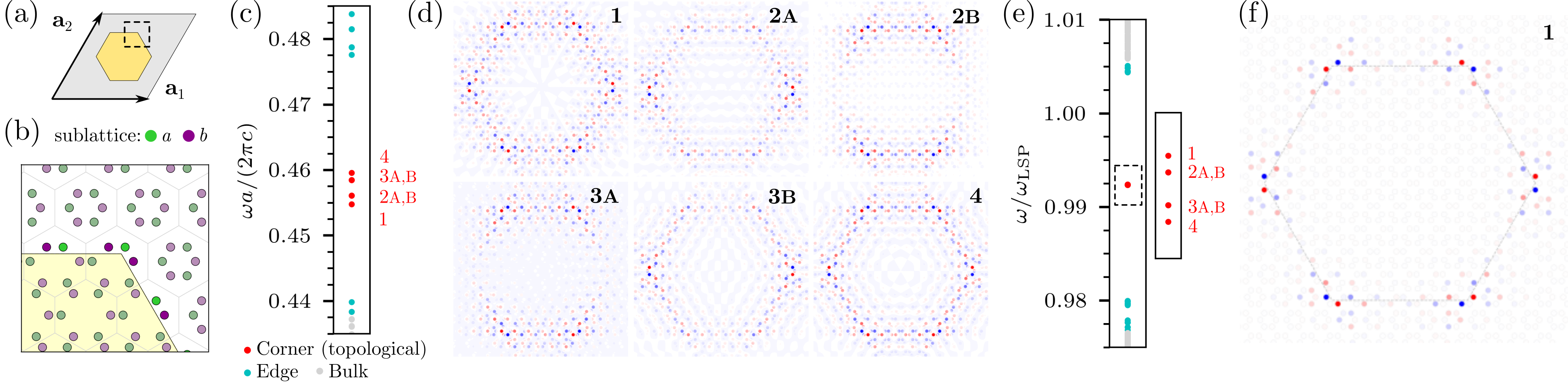}
    \caption{
    (a) Scheme for the topological particle supercell. (b) Particle lattice, with sublattices $a$ (green) and $b$ (purple). (c,d) Modes of a photonic crystal particle: Frequency ($\omega$) of topological corner (red), edge (cyan) and bulk (grey) states (c), and displacement field plots, showing $D_z$ (d). (e,f) Quasistatic model of the topological particle. (e) Frequency of topological corner, edge and bulk states, for silver nanoparticles with radius $10$~nm and height $40$~nm. (f) Dipole moments of the six corner eigenmodes. In the color scale used in (d) and (f) red (blue) represents positive (negative) values. In both cases, $\delta = 0.11$. }
    \label{fig:photonic_crystal_particle}
\end{figure*}

Figure~\ref{fig:photonic_crystal_bulk}(b) presents the band structure of the expanded phase for $\delta=0.11$. We first determine the topological properties of the system though the application of topological quantum chemistry~\cite{NaturePaper,dePaz2019}. 
The irreducible representations of the eigenfields at the high symmetry points (irrep labels), displayed in the band structure, are calculated using GTPack~\cite{gtpack1,gtpack2}. Using the catalogue of Elementary Band Representations (EBRs) in the Bilbao Crystallographic Server~\cite{Bilbao1,Bilbao2,Bilbao3, NaturePaper,GraphDataPaper,GroupTheoryPaper}, along with the irrep labels we can identify the topological properties of each set of connected bands of our PhC. 
Counting from $\omega = 0$, bands 4-6 are all interconnected and their irrep labels are accordant to Wannier functions centered in the $3c$ Wyckoff position transforming in the $(E_1\uparrow G)_{3c}$ band representation.
Since these bands can be identified with an EBR, we can conclude that the system  presents a trivial $\mathbb{Z}_2$ topological invariant. Nevertheless, the $3c$~Wyckoff position of the band representation indicates that the Wannier functions of this set of bands are not centered around the origin of the unit cell, but at their edges. This situation can be understood as a 2D analog of the topological hybridization of eigenstates of a 1D SSH chain. This topological phase was labeled in the past in analogy with solid-state systems as the photonic obstructed atomic limit (OAL), because although an atomic limit exists it is `obstructed' since the Wannier centers are not located at the position where the photonic “atoms” sit~\cite{dePaz2020}. Note that here the photonic atom is the collection of the six contracted/expanded cylinders inside the unit cell. Moreover, 
we characterize our system through the calculation of the eigenvalues of the Wilson loop~\cite{dePaz2020} for this set of connected bands, Fig.~\ref{fig:photonic_crystal_bulk}(c). 
The resulting Wilson loops present no windings (which are characteristic of $\mathbb{Z}_2$ or Chern insulators), but the Wannier centers are not only localized in the origin of the unit cell ($W = 0$) as in a trivial system, but also at its edges ($W = \pm \pi$), indicating that the system presents an obstruction similar to the 1D SSH chain~\cite{vanderbilt2018}. On the other hand, the PhC in the contracted phase is a trivial photonic insulator. This can be seen from the Wannier centers of the EBRs being located at the origin of the unit cell ($1a$~Wyckoff position), and by looking at the eigenvalues of Wilson loop
(see~\cite{supplemental}). 


It should be emphasized here that in 2D systems, there is a subtle relationship between OAL and HOTIs. In toy models with nearest neighbour interactions, it is often possible to define a chiral symmetry, which forces the spectrum to be symmetric about a fixed energy (often taken to be zero in the literature). If an OAL model has chiral symmetry, then it is sometimes possible to define a bulk topological invariant which counts the number of 0D corner modes in a finite-sized system preserving the crystal symmetries~\cite{miert2020topological}. Systems with non-zero values of this invariant are properly termed HOTIs. In the absence of chiral symmetry, as is the case in photonic systems with long-range interactions, however, there is no guarantee that a finite-sized system will have corner modes pinned to a special frequency. These systems are regarded as OAL systems, and can be characterized by the centers of their Wannier functions (as above), by real-space invariants 
(see~\cite{supplemental} and Ref.~\cite{song2020twisted}), or a filling anomaly (see ~\cite{supplemental} and Refs.~\cite{wieder2020strong,benalcazar2018quantization}). 
In order to make semi-analytical predictions about the presence and robustness of corner modes, for the remainder of this work we will exploit the fact that our model is deformable to a chiral-symmetric limit although this symmetry is strictly broken by unavoidable long-range interactions.



While the 0D corner modes in 2D SSH-like PhC particles (that is, finite size crystals containing several unit cells) with $C_4$ symmetry have been extensively explored~\cite{ota2018photonic,xie2019visualization,chen2019direct,kim2020lasing, han2020lasing}. In photonic crystal particles with $C_6$ symmetry, only the 1D edge states have been studied~\cite{Siroki2017, Jalali2020,barik2019}.
Firstly, we analyze the emergence of 0D photonic corner states in this system by looking at 2D particles made of cells in the expanded phase and surrounded by cells in the contracted phase, see Figs.~\ref{fig:photonic_crystal_particle}(a) and~\ref{fig:photonic_crystal_particle}(b)~\footnote{We build supercells of 21 unit cells in the $\mathbf{a}_1$ and $\mathbf{a}_2$ lattice directions, filling a central hexagonal portion of the supercell with 5 lattice constants in the expanded phase ($\delta=0.11$). To prevent the leaking of energy 
to the vacuum, we surround the central hexagon by cells in the contracted phase ($\delta=-0.11$), which behaves as a trivial photonic insulator with a matched band gap.}.
Results of MPB supercell calculations~\cite{johnson2001block} are shown in Figs.~\ref{fig:photonic_crystal_particle}(c) and~\ref{fig:photonic_crystal_particle}(d). The frequency eigenvalues ~\ref{fig:photonic_crystal_particle}(c) show a clear band-gap
, with 6 mid-gap states.
The real part of the displacement field eigenvectors $D_z$ for these 6 states are shown in Fig.~\ref{fig:photonic_crystal_particle}(d). These are concentrated at the corners of the particles, thus classify them as corner modes | marked in red in Fig.~\ref{fig:photonic_crystal_particle}(c). States $2$A,B and $3$A,B are degenerate pairs. The states immediately above and below the bandgap can be classified as edge states (cyan) \cite{Siroki2017,barik2019,Jalali2020}, followed by bulk eigenstates (gray). Thus, the 0D corner states in this PhC particle are hosted within the gapped 1D edge states, in contrast to HOTMs in $C_{3}$- and $C_4$-symmetric PhCs~\cite{xie2019visualization,chen2019direct,Li2020}. 

\textit{Coupled dipole model.|} Since the spectrum of the PhC particle is determined by lattice symmetries together with long-range interactions, we now exploit a semi-analytical model to unveil the properties of corner modes in a closely related nanophotonic system. The coupled dipole model is a versatile method for investigating the optical response of arrays of subwavelength elements such as cold atoms or plasmonic nanoparticles (NPs)~\cite{abajo2007colloquium}. Within this model we can reproduce all the relevant features found in full field simulations of the PhC topological particle. Then, we use it to shed further light on the properties of the corner modes, particularly on their robustness against disorder. This model goes beyond tight-binding, nearest neighbour models by including interactions between all the lattice elements (excluding self-interactions) with the appropriate propagator.
In this formalism, the modes can be found by solving a generalised eigenvalue equation, 
\begin{align}
    \left( \sum_{i\neq j} \hat{\mathbf{I}}\frac{1}{\alpha(\omega)} -  \hat{\textbf{G}}(\textbf{d}_{ij}, \omega)\right) \cdot \mathbf{p}_j = 0,
    \label{eqn:CDA}
\end{align}
where $\mathbf{p}_j$ are the dipole moments, $\hat{\textbf{G}}$ is the dyadic Green's function that describes dipole-dipole interactions, $\alpha$ is the polarizability of the subwavelength elements, $\omega$ is the frequency and the separation between NPs is $\mathbf{d}_{ij} = \mathbf{d}_i - \mathbf{d}_j$. 
The specifics of the physical dipolar elements enter through the  polarizability, from which the resonance frequencies of the modes can be extracted~\cite{supplemental}.
In Figs.~\ref{fig:photonic_crystal_particle}(e) and~\ref{fig:photonic_crystal_particle}(f) we present results of the dipole model for the OAL particle with the same geometry as the PhC in Figs.~\ref{fig:photonic_crystal_particle}(c) and~\ref{fig:photonic_crystal_particle}(d). Here we particularise the system to the out-of-plane modes of subwavelength spheroidal metallic NPs, which correspond to the TM modes in the PhC~\footnote{We take silver NPs with  parameters $\epsilon_\infty = 5$, $\omega_p = 8.9$~eV~\cite{Yang2015}, radius $r = 10$~nm and height $h = 40$~nm.}. We take a quasistatic approximation, and only include the near-field interaction term in the Green's function ($\propto 1/d^3$), which is accurate for these subwavelength NPs. In this approximation, the eigenvalues of Eq.~\eqref{eqn:CDA}, $E=1/\alpha(\omega)$, only depend on the particular geometrical arrangement of the dipoles. Figure~\ref{fig:photonic_crystal_particle}(e) shows the frequency spectrum around the band gap with corner modes within the gapped edge and bulk bands~\footnote{For plasmonic NPs the energy ordering of the modes is opposite to that of dielectric cylinders. 
This is because the bonding mode of out-of-plane dipoles which minimises energy corresponds to the hexapole, while the monopole has antibonding mode and lies at highest energy.}. 
For the plasmonic system, zero eigenvalue ($E=0$) maps to $\omega_\text{LSP}$, the localized surface plasmon frequency of the NPs. We see that the center of the band gap is located close to but not exactly at $\omega_\text{LSP}$, and that the spectrum is not exactly symmetric around that point. This is a consequence of chiral-symmetry breaking due to long-range interactions, as we discuss below in detail.


In Fig.~\ref{fig:photonic_crystal_particle}(f) we plot the real space dipole moments of the first mid-gap corner eigenmode, which reproduce well the $D_z$ field distributions of the PhC. 
Importantly, the corner modes are localized on a particular sublattice, while the dipole moments in the opposite sublattice remain virtually zero, shown in Fig.~\ref{fig:photonic_crystal_particle}(b). A similar sublattice localization of coner modes is present in the PhC, though weaker due to the fully retarded interactions. Nevertheless, this shows that both systems are approximately chiral-symmetric despite the long-range interactions, 
which has implications on the robustness of these 0D modes. 
In addition, these modes are well separated from the gapped bulk and edge states and are tightly confined to the corners.

\begin{figure}
    \includegraphics[width=\columnwidth]{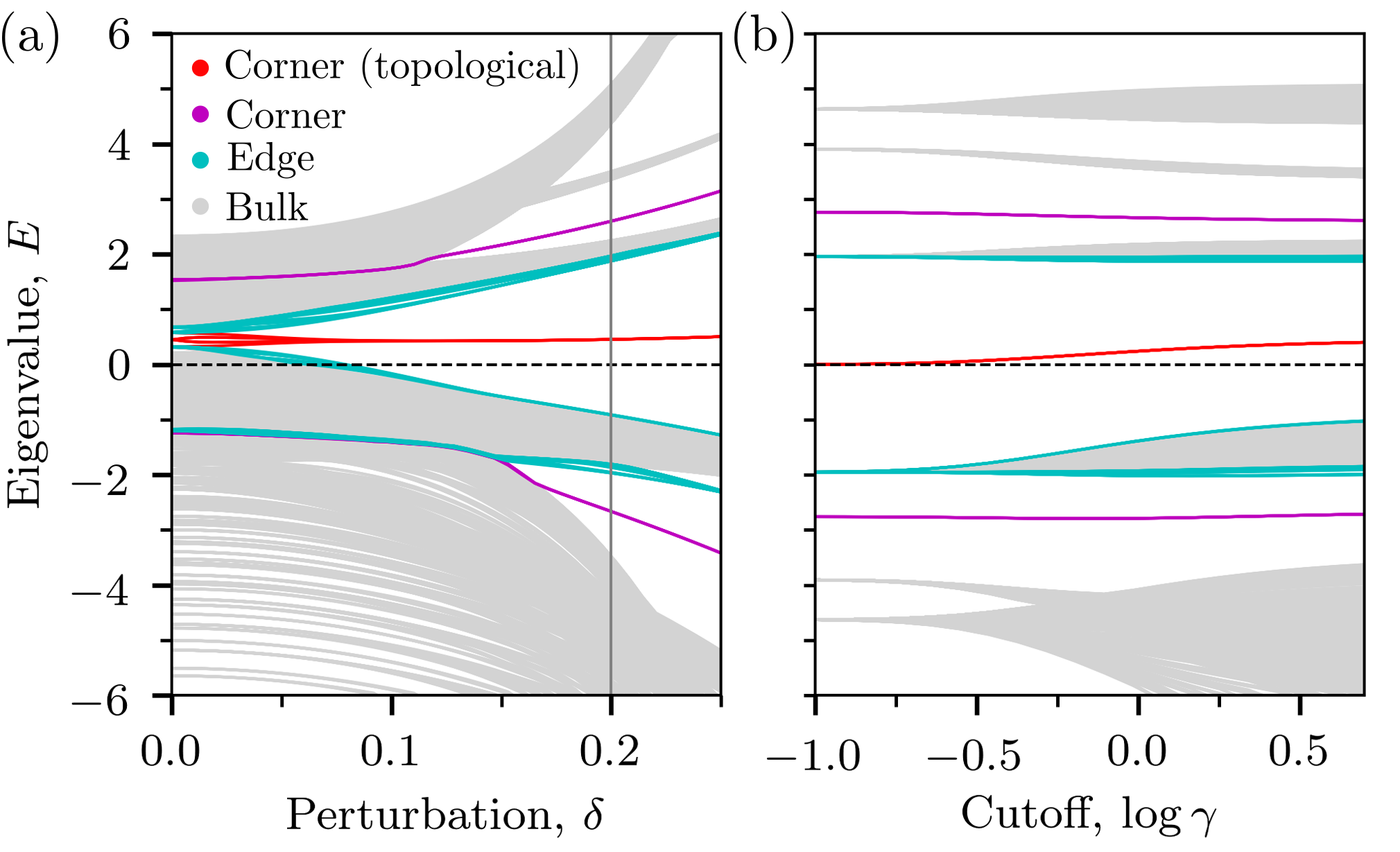}
    \caption{Topological particle eigenvalues. (a) Evolution with increasing unit cell perturbation $\delta$, with topological corner modes (red) well separated from the edge (cyan) and bulk (grey) modes. Other corner modes are shown in magenta. (b) Dependence on interaction length between lattice sites, $\gamma$, for $\delta = 0.2$. As interactions go from nearest neighbours $\gamma=0.1$ to long range, chiral symmetry is broken and the spectrum is no longer symmetrical about $E=0$.} 
    \label{fig:quasistatic}
\end{figure}

We now use the coupled dipole model to better characterise the properties of the corner modes. First, we study the behaviour of the system as a function of $\delta$, the deviation the lattice of NPs away from a perfect honeycomb. In Fig.~\ref{fig:quasistatic}(a), we plot the eigenvalue spectrum,
such that the symmetry properties of the spectrum around zero eigenvalue are clearer. 
Starting from the unperturbed honeycomb lattice  ($\delta=0$), we see how increasing $\delta$ controls the size of the bulk band-gap. At the same time, the corner modes (red) stay at approximately constant eigenvalue, only slightly shifted away from zero due to the inherent breaking of chiral symmetry.
In addition, edge modes (cyan) appear at the edges of the bulk bands.  As $\delta$ increases, the corner modes are more isolated in the band structure, and hence more strongly confined to corners of the particle. 
For $\delta \gtrsim 0.12$ new sets of corner modes (magenta) emerge from the bulk for positive and negative eigenvalues.
In contrast to the corner modes discussed here, these modes do not lie at the middle of the gap, and 
they are not localized only on one of the sublattices. 

The coupled dipole model also enables us to analyze the photonic corner modes analytically, as detailed in the SM~\cite{supplemental}. 
We find that when interactions are short-range, the eigenvalue problem for the coupled dipoles maps onto a tight-binding Schr\"odinger equation for a system with six $s$-orbitals at the $6d$~Wyckoff position in the unit cell (there is one $s$-orbital at the position of each NP). As $\delta$ increases, the model undergoes a transition between an atomic limit phase with Wannier centers on the $1a$ position, to an OAL phase with Wannier centers on the $3c$ position; in the short-range limit these Wannier functions are compactly supported, and can be found exactly. For a finite-sized system, the two atomic limits are distinguished by the $p6mm$ real space invariants of Ref.~\cite{song2020twisted}, which confirms that HOTMs are protected by lattice symmetries. Furthermore, we can solve for the corner modes in a topological particle in the long-wavelength approximation. We find that the low-energy theory of the domain between trivial and OAL particle naively resembles the edge of a quantum-spin Hall (QSH) insulator if only the lowest-order terms are considered.
However, when we include crystalline- and chiral-symmetric perturbations, we find that the QSH edge states gap to yield six corner modes pinned to mirror lines and related by sixfold rotational symmetry. Since the corner modes are eigenstates of the chiral symmetry, they must be localized to a single sublattice. We can then include chiral symmetry breaking perturbatively to find that the corner modes are lifted from zero eigenvalue (or $\omega = \omega_{\mathrm{LSP}}$), consistent with calculations as we discuss next. 

We study the effect of long-range interactions by introducing an artificial cut-off in the coupled dipole model. We introduce an exponential decay to the dipole-dipole interactions,
$f_{\mathrm{c.o.}}(d_{ij}) = \exp[-(d_{ij} - d_{ij}^0)/(d_{ij}^0\gamma)]$, where $d_{ij}^0$ is the nearest neighbour separation for each dipole and $\gamma$ is a cut-off parameter to control the interaction range~\cite{supplemental}. 
This allows us to continuously tune the interaction range from nearest neighbours ($\gamma = 0.1$), to electronic-like exponentially suppressed ones, all the way to full dipolar interactions ($\gamma \approx 5$), 
as we show in Fig.~\ref{fig:quasistatic}(b) for fixed $\delta = 0.2$. For small values of $\gamma$, interactions in practice are only between nearest neighbours, such that there is no coupling between dipoles of the same sublattice. 
This preserves chiral symmetry and results in a spectrum that is symmetric about zero eigenvalue, with six degenerate topological corner modes (red) that are pinned at zero. 

Increasing the range of the interaction breaks chiral symmetry through coupling of elements in the same sublattice. This shifts the corner modes away from zero eigenvalue, lifts their degeneracy (from six degenerate states to 1+2+2+1, as in Fig.~\ref{fig:photonic_crystal_particle}), and removes the symmetry of the spectrum about zero eigenvalue [or $\omega = \omega_{\mathrm{LSP}}$ in Fig.~\ref{fig:photonic_crystal_particle}(e)]. 
Finally, it is interesting to note that the other set of corner modes (magenta) are 
not pinned at zero even for nearest neighbour interactions. 
This is different from the type II corner states identified in Ref.~[\onlinecite{Li2020}] for the breathing kagome lattice, which emerge due to long-range interactions. 

\begin{figure}
    \centering
    \includegraphics[width=\columnwidth]{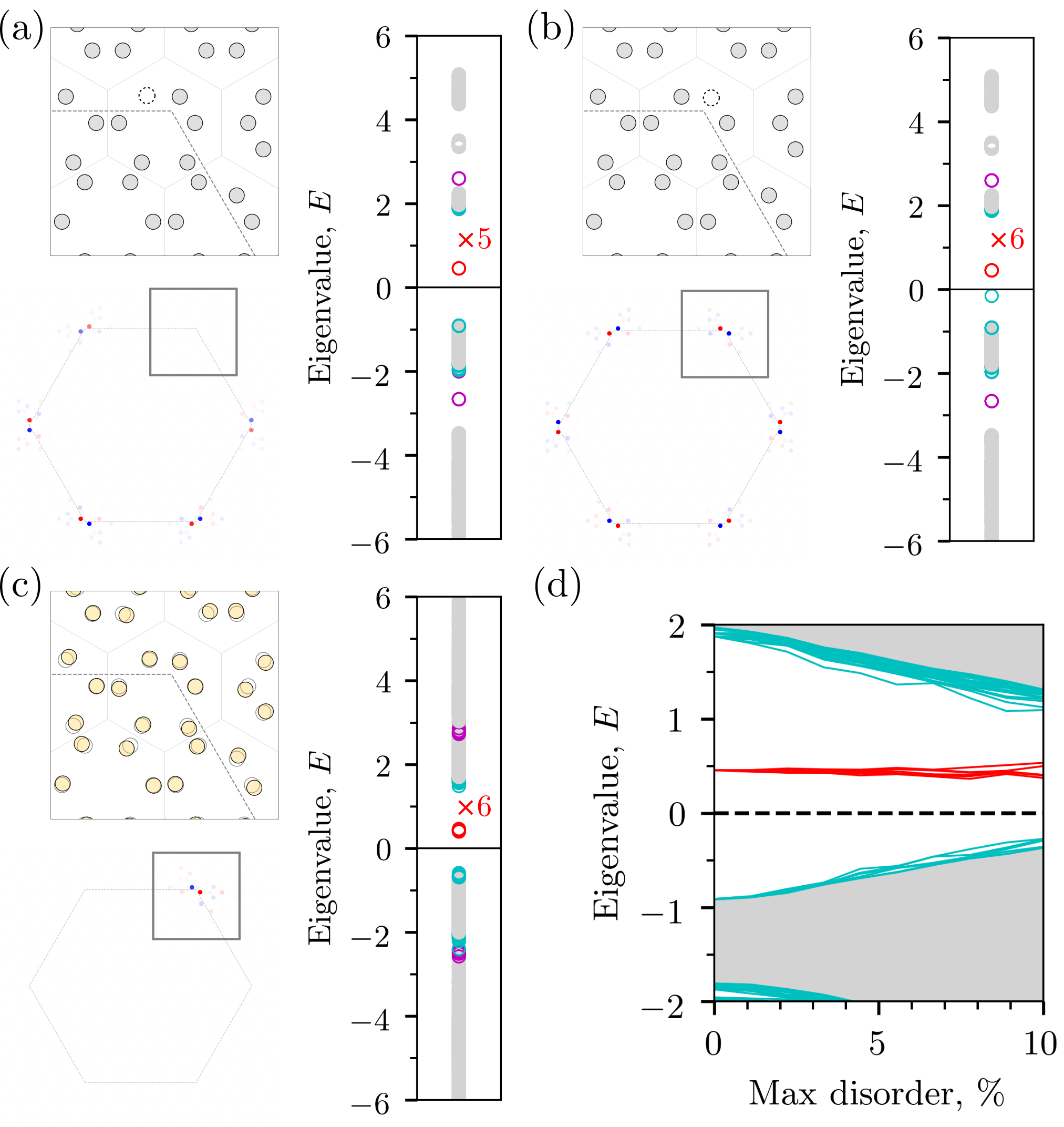}
    \caption{Robustness of corner states against defects and disorder in the quasistatic model, for $\delta=0.2$. (a) A $C_6$-symmetry breaking defect in the lattice affects the mid-gap corner modes. (b) The corner states are robust against another kind of $C_6$-symmetry breaking defect due to the corner modes being close to chiral-symmetric. (c) The degeneracy of the corner modes is lifted by random disorder: the position of the lattice sites is shifted randomly up to 5\%. (d) Eigenvalue spectrum for increasing positional disorder.}
    \label{fig:disorder}
\end{figure}

\textit{Robustness against defects and disorder.|} We now take advantage of the coupled dipole model to test the degree of protection of the corner modes against defects. 
Hence, we quantify protection by evaluating if the number of states within the band gap, together with the symmetries and degeneracies they satisfy, are left invariant. 
First, we create a strong defect in the crystal by removing one lattice site next to the corner of the particle,
Fig.~\ref{fig:disorder}(a). Since this breaks the $C_6$ and mirror symmetries that protect the corner modes, one of them disappears and the remaining five satisfy new symmetry relations and degeneracies, see field plots and eigenvalue spectrum in Fig.~\ref{fig:disorder}(a). Next, we consider removing one lattice site at exactly the corner Fig.~\ref{fig:disorder}(b), breaking the $C_6$ symmetry but respecting one mirror symmetry. Remarkably, the corner states are robust against this defect: there are 6 mid-gap states and they satisfy the same symmetries and degeneracies as before the perturbation. 
This is a consequence of the system being deformable to a chirally-symmetric system. Despite the presence of long-range interactions, the modes still sit on alternate sublattices, and the mode intensity is virtually zero at the removed lattice site. 

Finally, we test robustness against random positional disorder. In Fig.~\ref{fig:disorder}(c) we consider a system with maximum $5\%$ random disorder in lattice sites. Crucially, this breaks the $C_6$ symmetry across the whole lattice, such that the degeneracies of the corner modes are lifted, and each of the six mid-gap states localizes at one of the corners. On the other hand, we see in the spectrum how, despite the other corner modes and edge modes being lost to the bulk,
the mid-gap corner modes remain well isolated at mid-gap energies. For practical purposes they are robust against random spatial perturbations. This is confirmed in Fig.~\ref{fig:disorder}(d), where
we plot a close up of the band gap and the HOTMs for increasing random positional disorder, up to a maximum of $10\%$. 

\textit{Conclusions.|}
We have studied the emergence of topologically protected corner modes in breathing honeycomb PhC particles. By analyzing the lattice through topological quantum chemistry, Wilson loops and the calculation of real space topological invariants, we conclude that  the topological properties emerge from an obstructed atomic limit phase, which in 2D is reminiscent of higher-order topology. Finally, we quantify the robustness of topological corner modes in PhCs to different kinds of perturbations. We conclude that, while long-range interactions inevitably break chiral symmetry, the corner modes are still protected by lattice symmetries. Although we have focused here on the breathing honeycomb lattice PhC, our analysis applies to all classical wave systems.

\smallskip
\begin{acknowledgments}
M.P. and P.A.H. acknowledge funding from the Leverhulme Trust. P.A.H. acknowledges funding from Funda\c c\~ao para a Ci\^encia e a Tecnologia and Instituto de Telecomunica\c c\~oes under projects CEECIND/03866/2017 and UID/EEA/50008/2020. B.B. acknowledges support of the Alfred P. Sloan foundation.  M.G.V. acknowledges support from DFG INCIEN2019-000356 from Gipuzkoako Foru Aldundia and the Spanish Ministerio de Ciencia e Innovacion (grant number PID2019-109905GB-C21). D.B. acknowledges supported by the Spanish Ministerio de Ciencia, Innovation y Universidades (MICINN) through the project FIS2017-82804-P, and by the Transnational Common Laboratory \emph{Quantum-ChemPhys}.
\end{acknowledgments}



\bibliography{main.bib}

\end{document}


\title{Supplemental Material:\\ On the robustness of topological corner modes in photonic crystals}

\author{Matthew Proctor}
\affiliation{%
Department of Mathematics, Imperial College London, London, SW7 2AZ, U.K.}
\author{Paloma Arroyo Huidobro}%
 \email{p.arroyo-huidobro@lx.it.pt}
\affiliation{%
Instituto de Telecomunica\c c\~oes, Instituto Superior Tecnico-University of Lisbon, Portugal}%

\author{Barry Bradlyn}%
\email{bbradlyn@illinois.edu}
\affiliation{Department of Physics and Institute for Condensed Matter Theory, University of Illinois at Urbana-Champaign, Urbana, IL, 61801-3080, USA}%

\author{Mar\'{i}a Blanco de Paz}
\affiliation{Donostia International Physics Center, 20018 Donostia-San Sebasti\'an, Spain}
\author{Maia G. Vergniory}
\affiliation{Donostia International Physics Center, 20018 Donostia-San Sebasti\'an, Spain}
\affiliation{IKERBASQUE, Basque Foundation for Science, Maria Diaz de Haro 3, 48013 Bilbao, Spain}
\author{Dario Bercioux}
\affiliation{Donostia International Physics Center, 20018 Donostia-San Sebasti\'an, Spain}
\affiliation{IKERBASQUE, Basque Foundation for Science, Maria Diaz de Haro 3, 48013 Bilbao, Spain}
\author{Aitzol Garc\'{i}a-Etxarri}
\email{aitzolgarcia@dipc.org}
\affiliation{Donostia International Physics Center, 20018 Donostia-San Sebasti\'an, Spain}
\affiliation{IKERBASQUE, Basque Foundation for Science, Maria Diaz de Haro 3, 48013 Bilbao, Spain}
\maketitle


\section{Bulk Band Structures and Wilson Loops}

Here we present the unit cell arrangements, as well as the band structure and the Wilson loops characterization for the lowest bands of the expanded and contracted lattice.
The Wilson loops for the expanded lattice show that the Wannier functions are centered at the 3c position, whereas for the contracted lattice they are centered at 1a. 

\begin{figure}[h]
    \centering
    \includegraphics[width=\linewidth]{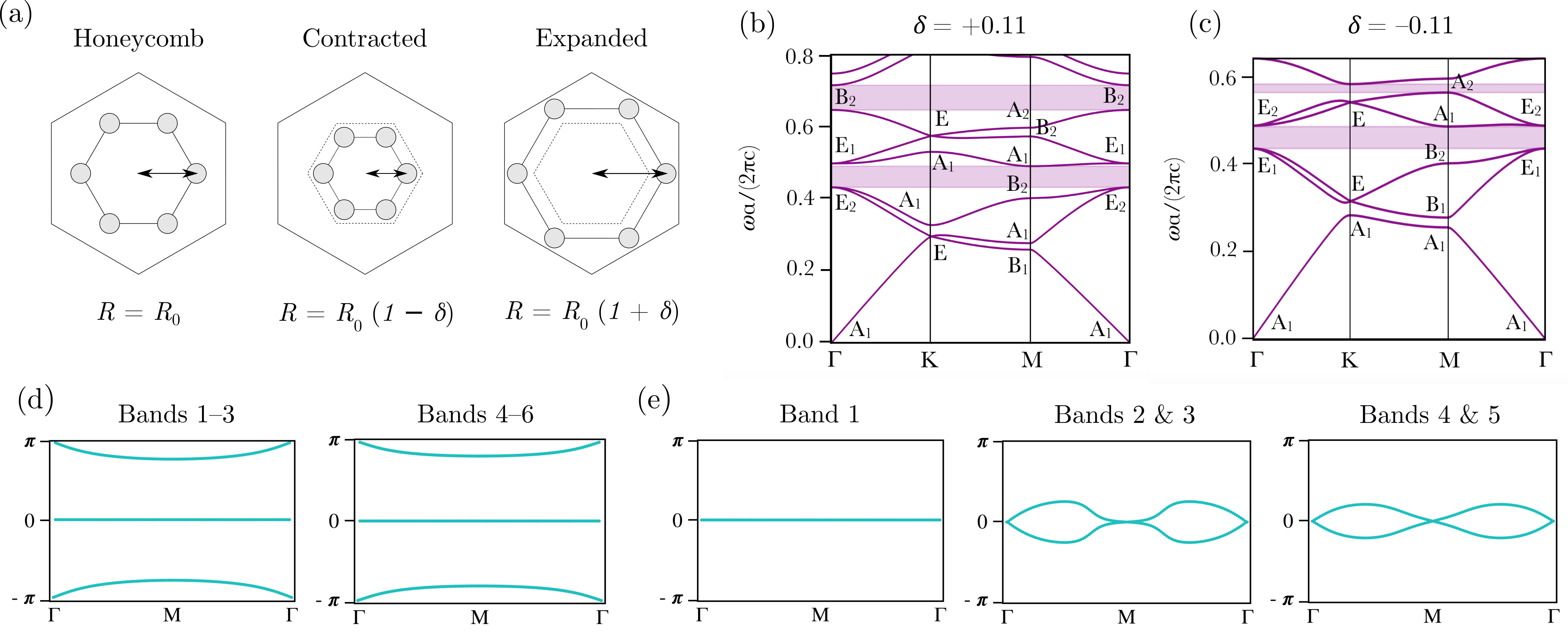}
    \caption{(a) Unit cell arrangements for the honeycomb, contracted and expanded lattices. Band structures and Wilson loops for  expanded lattice, $R = R_0 + 0.11$, in panels (b, d) and  contracted lattice, $R = R_0-0.11$, in panels (c, e).}
    \label{fig:bands_irreps_wl_exp_cont}
\end{figure}

\section{Coupled Dipole Model}

In the quasistatic (QS) approximation, we consider an array of point dipoles and model interactions between the them using the coupled dipole method \cite{abajo2007colloquium}. In the absence of an external electric field, the (electric) dipole moment at position $\mathbf{d}_i$ due to a dipole at position $\mathbf{d}_j$ is given by,
\begin{align}\label{eqn:cda}
    \frac{1}{\alpha(\omega)}\mathbf{p}_i = \hat{\textbf{G}}(\textbf{d}_{ij}, \omega) \cdot \mathbf{p}_j,
\end{align}
where $\omega$ is the frequency and the separation between dipoles is $\mathbf{d}_{ij} = \mathbf{d}_i - \mathbf{d}_j$. The dyadic Green's function, which describes the dipole-dipole interaction; it can be written as:
\begin{align}\label{eqn:dyadic_gf}
    \hat{\textbf{G}}(\textbf{d}_{ij}, \omega) &= k^2\frac{e^{ikd}}{d} \biggl[
    \biggl(
    1 + \frac{i}{kd} - \frac{1}{k^2d^2}
    \biggr)\hat{\textbf{I}} \,- 
    \biggl(
    1 + \frac{3i}{kd} - \frac{3}{k^2d^2}
    \biggr)\textbf{n}\otimes\textbf{n}
    \biggr],
\end{align}
where $d = |\mathbf{d}_{ij}|$, $\mathbf{n}=\mathbf{d}_{ij}/d$ and wavenumber $k = \sqrt{\epsilon_m}\omega/c$; we assume the permittivity of the medium $\epsilon_m = 1$. 
In the QS approximation, we retain only the quickly decaying $1/d^3$ terms in the Green's function by letting $k\rightarrow0$. Then for a periodic array of dipoles, we can we write the following eigenvalue equation, 
\begin{align}\label{eqn:eigenvalue}
\left(\hat{\mathbf{I}}\frac{1}{\alpha(\omega)} - \hat{\textbf{H}}(\textbf{k}_B, \omega)\right)\cdot\textbf{p} = 0, 
\end{align}
where $\mathbf{p}$ is a vector which contains all dipole moments in the unit cell. The interaction matrix $\hat{\textbf{H}}(\textbf{k}_B, \omega)$ has elements,
\begin{align}\label{eqn:interaction_matrix}
    H_{ij}=
    \begin{cases}
    \sum\limits_{\textbf{R}} \hat{\textbf{G}}(\textbf{d}_i - \textbf{d}_j + \textbf{R}, \omega) \hspace{2px} e^{i\textbf{k}_B\cdot\textbf{R}} & i \neq j\\
    \sum\limits_{|\textbf{R}|\neq0} \hat{\textbf{G}}(\textbf{R}, \omega) \hspace{2px} e^{i\textbf{k}_B\cdot\textbf{R}} & i = j
    \end{cases},
\end{align}
with Bloch wavevector $\mathbf{k}_B$ and lattice sites $\mathbf{R} = n\mathbf{a}_1 + m\mathbf{a}_2$, where the lattice vectors are defined in the main text.
The dipole model accurately describes a nanophotonic system of resonators such as metallic nanoparticles (NPs), provided the NP radius satisfies $r<3R$, where $R$ is the nearest neighbour spacing. The optical response of an individual NP is given by the polarizability $\alpha(\omega)$. In the following, we assume a static polarizability,
\begin{align}
    \alpha(\omega) = \frac{V}{4\pi}\frac{\epsilon(\omega) - 1}{L\,[\epsilon(\omega) + 2]},
\end{align}
$V$ is the NP volume, $L$ is a geometrical factor and $\epsilon(\omega)$ is the Drude permittivity \cite{Moroz2009}. The quasistatic Drude permittivity is written,
\begin{align}
    \epsilon(\omega) = \epsilon_\infty  - \frac{\omega_p^2}{\omega^2}.
\end{align}
In this manuscript, we use silver spheroidal NPs with material parameters $\epsilon_\infty = 5$, $\omega_p = 8.9$~eV and size parameters radius $r = 10$~nm, height $h = 40$~nm.~\cite{Yang2015} The spheroidal shape causes the in-plane and out-of-plane resonances of the NP to split in frequency and become completely decoupled, meaning we can consider them separately. To make comparisons with the 2D photonic crystal, we only consider the out-of-plane interactions and take the $\hat{z}\hat{z}$ component of the dyadic in Eq.~\eqref{eqn:dyadic_gf}, $\hat{\textbf{G}}(\textbf{d}_{ij}, \omega) = -1/r^3$. The size of the interaction matrix in \label{eqn:interaction_matrix} will then be $N\times N$ where N is the number of elements in the supercell.
Additionally, to model a finite system we only consider normal incidence and solve the eigenvalue problem at $\Gamma$, $\mathbf{k}_B = (0, 0)$.

\section{Topological analysis of the quasistatic model: Real Space Invariants}
To analyze the topological properties of our nanophotonic resonator (coupled dipole) system, we can reinterpret the interaction matrix $H_{ij}$ of the quasistatic model as a (long-ranged) Hamiltonian for a topological phase transition. While $H_{ij}$ is in general long range (it has power-law decaying matrix elements in position space) which can lead to cusp singularities in the band structure (which are removed when a fully retarded Green's function is used), we can nevertheless probe the presence and topological protection of edge and corner modes originating from analytic regions in the band structure. To this end, we can truncate the interaction matrix at the nearest neighbor level. Doing so, we can reinterpret $H_{ij}$ as a tight-binding model for dipolar resonators at the $6d$ Wyckoff position in space group $p6mm$. In reduced coordinates, the positions of the dipoles are $q_0=(s,0),q_1=(s,-s),q_2=(0,-s),q_3=(-s,0),q_4=(-s,s),q_5=(0,s)$. In the basis of these six orbitals, the $C_6$ symmetry is represented by
\begin{equation}
    \rho(C_6)=\left(\begin{array}{cccccc}
    0 & 1 & 0 & 0 & 0 & 0 \\
    0 & 0 & 1 & 0 & 0 & 0 \\
    0 & 0 & 0 & 1 & 0 & 0 \\
    0 & 0 & 0 & 0 & 1 & 0 \\
    0 & 0 & 0 & 0 & 0 & 1 \\
    1 & 0 & 0 & 0 & 0 & 0
    \end{array}\right),
\end{equation}
mirror symmetry about the $x$-axis is represented by
\begin{equation}
    \rho(m_x)=\left(\begin{array}{cccccc}
    1 & 0 & 0 & 0 & 0 & 0 \\
    0 & 0 & 0 & 0 & 0 & 1 \\
    0 & 0 & 0 & 0 & 1 & 0 \\
    0 & 0 & 0 & 1 & 0 & 0 \\
    0 & 0 & 1 & 0 & 0 & 0 \\
    0 & 1 & 0 & 0 & 0 & 0
    \end{array}\right),
\end{equation}
and time-reversal symmetry is represented by complex conjugation. We can write the interaction matrix $H_{ij}$ as the sum of two terms
\begin{subequations}
\begin{equation}
    H(\mathbf{k},s)=(1-t(s))M + t(s) N(\mathbf{k})
\end{equation}
where
\begin{equation}
    M=\left(\begin{array}{cccccc}
    0 & 1 & 0 & 0 & 0 & 1 \\
    1 & 0 & 1 & 0 & 0 & 0 \\
    0 & 1 & 0 & 1 & 0 & 0 \\
    0 & 0 & 1 & 0 & 1 & 0 \\
    0 & 0 & 0 & 1 & 0 & 1 \\
    1 & 0 & 0 & 0 & 1 & 0
    \end{array}\right),
\end{equation}
\begin{equation}
    N(\mathbf{k})  = \left(\begin{array}{cccccc}
    0 & 0 & 0 & e^{ik_1} & 0 & 0\\
    0 & 0 & 0 & 0 & e^{i(k_1-k_2)} & 0 \\
    0 & 0 & 0 & 0 & 0 & e^{-ik_2} \\
    e^{-ik_1} & 0 & 0 & 0 & 0 & 0 \\
    0 & e^{-i(k_1-k_2)} & 0 & 0 & 0 & 0\\
    0 & 0 & e^{ik_2} & 0 & 0 & 0
    \end{array}\right)
\end{equation}
\end{subequations}
Here $M$ is the intra-cell hopping matrix, and $N$ is the inter-cell hopping matrix. Note that this is written in an embedding where we keep the positions of the dipoles fixed at $s=0$, while we vary the hoppings $t(s)$ in accordance with the analysis of the main text (c.f. the treatment of the Peierls transition in the Su-Schrieffer-Heeger model~[\onlinecite{ssh1979}]). The function $t(s)$ smoothly and monotonically interpolates between $t(0)=0$ in the maximally contracted (triangular) lattice, and $t(1)=1$ in the maximally expanded (kagome) lattice. There is a critical point $t(s_*)=1/2$, where the intra- and inter-cell hopping amplitudes are equal. $H(\mathbf{k},s)$ has a gap at zero energy for all $\mathbf{k}$ and all $s\neq s_*$, with three negative and three positive energy bands. 

We will now proceed to show that the critical point separates the trivial and obstructed atomic limit phases of our model. First, we will compute the band representations carried by the occupied (negative energy) states in both gapped phases, and show that there is a transition between a phase with Wannier centers at the $1a$ position, and a phase with Wannier centers at the $3c$ position. Furthermore, we will show that the little group representations in these phases are consistent with what is found in the photonic crystal model. Then we will compute the ``real space invariants\cite{song2020twisted}'' for the trivial and OAL phases, and show that point group symmetric topological particles in the two phases are topologically distinct. Finally, by analyzing the low-energy theory of the critical point $H(\mathbf{k},s_*+\delta s)$ we will show that the interface between the trivial and topological phase must host a set of six corner states of topological origin.

\subsection{Band representation analysis}

Here we will establish that the Hamiltonians $H(\mathbf{k},s<s_*)$ and $H(\mathbf{k},s>s_*)$ describe topologically distinct atomic limits. To do so, let us note that, since $H(\mathbf{k},s)$ is gapped for all $s\neq s_*$, we can always adiabatically deform the Hamiltonian either to $s=0$ or $s=1$. It is thus sufficient to determine the topology of the bands when $s=0,1$.

Let us focus first on $s=0$, where we have
\begin{equation}
    H(\mathbf{k},0)=M.
\end{equation}
We can easily diagonalize the $\mathbf{k}$-independent matrix to find that the three occupied $(E<0)$ states have eigenvectors
\begin{subequations}
\begin{align}
    \mathbf{v}_1&=\frac{1}{\sqrt{6}}(1,-1,1,-1,1,-1)^T, \\
    \mathbf{v}_2&=\frac{1}{\sqrt{12}}(2,-1,-1,2,-1,-1)^T, \\
    \mathbf{v}_3&=\frac{1}{2}(0,1,-1,0,1,-1)^T, 
\end{align}
\end{subequations}
with corresponding energies
\begin{equation}
    E_1=-2,\;\; E_2=E_3=-1
\end{equation}
Since these eigenvectors give us $\mathbf{k}$-independent linear combinations of our basis orbitals, they can be Fourier transformed to yield exponentially localized (in fact, delta-function localized) Wannier functions at the $1a$ position of the unit cell. To determine the band representation under which these Wannier functions transform, we project the symmetry operations into the space of occupied states to obtain the sewing matrices
\begin{subequations}
\begin{align}
    \mathcal{B}^{(0)}(C_6)_{ij}\equiv \langle \mathbf{v}_i | \rho(C_6) | \mathbf{v}_j\rangle &= \begin{pmatrix} -1 & 0 & 0 \\ 0 & -\frac{1}{2} & \frac{\sqrt{3}}{2} \\ 0 & \frac{\sqrt{3}}{2} & -\frac{1}{2} \end{pmatrix}, \\
    \mathcal{B}^{(0)}(m_x)_{ij}\equiv \langle \mathbf{v}_i | \rho(m_x) | \mathbf{v}_j\rangle &= \begin{pmatrix}
    1 & 0 & 0 \\
    0 & 1 & 0 \\
    0 & 0 & -1
    \end{pmatrix}.
\end{align}
\end{subequations}
Comparing with the character tables on the Bilbao Crystallographic server, we see that this is the $B_2\oplus E_2$ representation of the site-symmetry group $G_{1a}\approx \mathrm{p}6\mathrm{mm}$ of the $1a$ Wyckoff position. Hence, when $s=0$ the occupied bands transform in the $(B_2\oplus E_2)_{1a}\uparrow G$ band representation\cite{NaturePaper,EBRTheoryPaper,GroupTheoryPaper}.

Next, let us analyze the case when $s=1$, where the Hamiltonian takes the form
\begin{equation}
    H(\mathbf{k},1)=N(\mathbf{k})
\end{equation}
We can diagonalize $N(\mathbf{k})$ to obtain the three occupied-band eigenvectors, which now have energies $E_1=E_2=E_3=-1$,
\begin{subequations}
\begin{align}
    \mathbf{w}_1&=\frac{1}{\sqrt{2}}(0,0,-e^{-ik_2},0,0,1)^T \\
    \mathbf{w}_2&=\frac{1}{\sqrt{2}}(0,-e^{i(k_1+k_2)},0,0,1,0)^T \\
    \mathbf{w}_3&=\frac{1}{\sqrt{2}}(-e^{ik_1},0,0,1,0,0)^T
\end{align}
\end{subequations}
Although these eigenvectors are $\mathbf{k}$-dependent, they are periodic and analytic, and hence can be Fourier transformed to yield compactly-supported Wannier functions. In this case, we can see from computing the position matrix elements
\begin{equation}
    \langle \mathbf{w}_i | \mathbf{x} | \mathbf{w}_j \rangle = -i \langle \mathbf{w}_i | \nabla_\mathbf{k} \mathbf{w}_j \rangle
\end{equation}
that these Wannier functions will be centered at the $3c$ Wyckoff position, with reduced coordinates $(1/2,0)$, $(0, 1/2)$, $(1/2,1/2)$. To determine under which band representation these Wannier functions transform, we can again compute the sewing matrices for the symmetry operations, yielding
\begin{subequations}
\begin{align}
    \mathcal{B}^{(1)}(C_6)_{ij}&\equiv \langle \mathbf{w}_i(C_6\mathbf{k}) | \rho(C_6) | \mathbf{w}_j(\mathbf{k})\rangle =\frac{1}{2} \begin{pmatrix} 
    0 & 0 & -e^{ik_1} \\ 
    1 & 0 & 0 \\ 
    0 & 1 & 0 
    \end{pmatrix}, \\
    \mathcal{B}^{(1)}(m_x)_{ij}&\equiv \langle \mathbf{w}_i(m_x\mathbf{k}) | \rho(m_x) | \mathbf{w}_j(\mathbf{k})\rangle = \begin{pmatrix}
    0 & -e^{i(k_1-k_2}) & 0 \\
    -e^{-ik_2} & 0 & 0 \\
    0 & 0 & 1
    \end{pmatrix}.  
\end{align}
\end{subequations}
Specializing to the high-symmetry points, we can verify that these are the sewing matrices obtained via induction from the $B_1$ representation of the site symmetry group $G_{3c}\approx \mathrm{p}2\mathrm{mm}$ of the $3c$ Wyckoff position. Thus, when $s=1$ the occupied bands transform in the $(B_1)_{3c}\uparrow G$ band representation. Thus, we have verified that as the parameter $s$ is tuned, the Hamiltonian $H(\mathbf{k},s)$ describes an obstructed atomic limit transition between the $1a$ and $3c$ Wyckoff positions. 

\subsection{Real Space Invariants}

Having established the presence of a bulk OAL transition for the Hamiltonian $H(\mathbf{k},s)$, we know that bulk systems with $s<s_*$ are topologically distinct from bulk systems with $s>s_*$. We would like to extend this analysis, however, to the case of finite-sized topological particles, and hence establish that the topological particles for the two different bulk phases are topologically distinguishable. To do this, we will employ the method of Real Space Invariants (RSIs) presented in Ref.~\cite{song2020twisted}. In that work, it was shown that there exist point group invariants which distinguish the classes of occupied states of a topological particle that can be deformed into each other through point group symmetric deformations of the Hamiltonian, as well as point-group symmetric addition of states from outside the topological particle. While these invariants are most generally formulated in terms of real-space point group irreps, in many cases they can be calculated from the momentum-space irreps of a band structure. In p6mm, there are seven invariants which can be computed in terms of the multiplicities of momentum-space irreps: they are
\begin{subequations}
\begin{align}
    \delta_{1,1a}&=n(M_3)-n(K_1)-n(\Gamma_2) \\
    \delta_{2,1a}&=n(\Gamma_3)+n(\Gamma_5)-n(\Gamma_2)-n(K_1) \\
    \delta_{3,1a}&=n(\Gamma_3)-2n(\Gamma_2)-n(\Gamma_6)-n(K_1)+n(K_2)+n(M_3) \\
    \delta_{1,2b}&=n(K_1)-n(\Gamma_1)-n(\Gamma_3) \\
    \delta_{1,3c}&=n(\Gamma_3)+n(\Gamma_6)-n(M_3) \\
    \delta_{1,6d}&=n(K_2)=n(K_1) \\
    \delta_{1,6e}&=2n(\Gamma_2)-2n(\Gamma_1)+n(K_1)-n(K_2)
\end{align}
\end{subequations}
where $n(\rho)$ is the multiplicity of the little group representation $\rho$ in the set of occupied bands. Note that each RSI is labelled by a Wyckoff position, indicating that it is an invariant computed from the set of orbitals localized to that Wyckoff position in the topological particle. 

For the case at hand, as $s$ is tuned from $0$ to $1$, our Hamiltonian undergoes a band inversion at the $\Gamma$ point. From the sewing matrices computed above, we find that as we tune from the trivial to the OAL phase, $n(\Gamma_5)$ decreases by $1$, while $n(\Gamma_6)$ increases by $1$. This implies that the real space invariants $\delta_{2,1a}, \delta_{3,1a},$ and $(-)\delta_{1,3c}$ each differ by $(-)1$ between the trivial and the OAL phases. This implies that even in a finite-sized topological particle, the trivial and OAL phases can be distinguished by their transformation properties under the point group 6mm. We now analyze the consequences of this distinguishability in terms of corner states.

\subsection{Corner states}

The key consequence of the topological distinction between the trivial and obstructed topological particles is the presence of protected corner states at a point-group symmetric boundary between the two phases. To see that the corner states are an inevitable consequence of the bulk topology, we will here adapt the method of Ref.~\cite{wieder2020strong} to analyze the low-energy theory of the topological particle system. This does not alter the topological properties of the Hamiltonian, but will simplify the analysis below. To begin, we replace the matrix $M$ with the spectrally flattened
\begin{equation}
    \tilde{M}=\mathbb{I}-2\sum_{i=1}^{3}\mathbf{v}_i\otimes\mathbf{v}_i,
\end{equation}
which shares the same negative energy eigenspace as the matrix $M$, but moves all states to the same eigenvalue $E_1=E_2=E_3=1$. We can then focus on the deformed Hamiltonian
\begin{equation}
    \tilde{H}(\mathbf{k},s)= (1-t(s))\tilde{M}+t(s)N(\mathbf{k})
\end{equation}
Our strategy here is to expand the Hamiltonian about the band-inversion point $(\Gamma,s_*)$, Fourier transform to position space, and allow the mass parameter $s$ to be spatially varying with $s(R)=s_*$, where $R\gg 1$. We will then perform a Jackiw-Rebbi analysis of the boundary states near $\mathbf{r}\approx R$, and analyze their stability to perturbations of the bulk Hamiltonian. Following this procedure, we will establish the existence of corner modes and a filling anomaly for our OAL topological particles even in the absence of chiral symmetry.

Let us project the Hamiltonian near $\mathbf{k}=0$, $s=s_*$ into the low-energy subspace of the topological band inversion. We find that at the gap-closing point, there is a fourfold band degeneracy at the $\Gamma$ point. This fourfold degenracy is the critical point between the trivial and OAL phases. Diagonalizing the critical Hamiltonian $\tilde{H}(0,s_*)=\tilde{M}+N(0)$ at the $\Gamma$ point, we find that the space of states at the critical point is spanned by the four zero-energy eigenvectors
\begin{align}
    \mathbf{u}_1&=\frac{1}{\sqrt{2}}(0,-1,0,0,0,1)^T, \\
    \mathbf{u}_2&=\frac{1}{\sqrt{2}}(-1,0,0,0,1,0)^T, \\
    \mathbf{u}_3&=\frac{1}{\sqrt{6}}(0,-1,0,2,0,-1)^T, \\
    \mathbf{u}_4&=\frac{1}{\sqrt{6}}(-1,0,2,0,-1,0)^T
\end{align}

After a suitable transformation to Cartesian coordinates, we can expand the Hamiltonian to first order in $\mathbf{k}$ and $m=t(s)-t(s_*)$ to find the Dirac-like Hamiltonian
\begin{equation}
    \tilde{H}(\mathbf{k},\delta s)\approx \frac{1}{4} (k_x\Gamma_x - k_y\Gamma_y -8m\Gamma_z) \label{eq:criticalham}
\end{equation}
where we have introduced anticommuting $4\times 4$ gamma matrices
\begin{subequations}
\begin{align}
    \Gamma_x&=\frac{1}{2}(\tau_z-\sqrt{3}\tau_x)\sigma_y=\sigma_y\tau_z', \\
    \Gamma_y&=\frac{1}{2}(\tau_x+\sqrt{3}\tau_z)\sigma_y=\sigma_y\tau_x',\\
    \Gamma_z&=\frac{1}{2}(\sigma_x\tau_0+\sqrt{3}\sigma_y\tau_y) \\
    \Gamma_4&=\frac{1}{2}(\sigma_y\tau_y - \sqrt{3}\sigma_x\tau_0)
\end{align}
\end{subequations}
where the $\bm{\tau}$ Pauli matrices act in  the block subspace of $\{(\mathbf{u}_1,\mathbf{u}_2),(\mathbf{u_3},\mathbf{u}_4)\}$, while the $\bm{\sigma}$ Pauli matrices act within the blocks.

We will now let $m\rightarrow m(\mathbf{r})$ depend on position. To be concrete, we assume that $m(\mathbf{r}\rightarrow 0) = -t_0, m(\mathbf{r}\rightarrow\infty) = t_0, m(\mathbf{r}=\mathbf{R})=0$, and we furthermore assume that  $m(\mathbf{r})=m(r)$ is circularly symmetric. We will look for zero-energy states localized near the domain wall $r=R$ by solving the eigenvalue equation~\cite{jackiw1976solitons}:
\begin{equation}
    (-i\partial_x\Gamma_x + i \partial_y\Gamma_y - 2m(r)\Gamma_z) f(r)|\phi\rangle =  Ef(r)|\phi\rangle
\end{equation}
Re-expressing this in polar coordinates, we have
\begin{align}
    \left[-2m(r)\Gamma_z -i\sigma_y\tau_1(\theta)\partial_r +i\sigma_y\frac{1}{r}\tau_2(\theta)\partial_\theta\right]f(r)|\phi\rangle = Ef(r)|\phi\rangle
\end{align}
where we have introduced
\begin{subequations}
\begin{align}
    \tau_1(\theta)&=\tau_z'\cos\theta -\tau_x'\sin\theta, \\
    \tau_2(\theta)&=\tau_z'\sin\theta + \tau_x'\cos\theta
\end{align}
\end{subequations}
We would like to look for solutions to this equation near $r=R$, where the mass changes sign. for $R$ sufficiently large, we can then treat the angular dispersion term $1/r\partial_\theta\approx 1/R\partial_\theta$ as a small perturbation. We will then find the spectrum of edge states by first solving
\begin{equation}
    \left[-2m(r)\Gamma_z -i\sigma_y\tau_1(\theta)\partial_r \right]f(r)|\phi\rangle = 0,\label{eq:radial}
\end{equation}
from which we will derive a low-energy edge Hamiltonian by projecting the angular velocity into this eigenbasis. Equation~\eqref{eq:radial} is solved by functions of the form
\begin{align}
    f(r)&\propto e^{-\int_R^r 2m(r') dr'} 
    -i\Gamma_z\sigma_y\tau_1(
    \theta)|\phi_i\rangle  -|\phi_i\rangle
\end{align}
We can write $|\phi_1\rangle, |\phi_2\rangle$ explicitly as
\begin{align}
    |\phi_1\rangle&=\frac{e^{i\theta/2}}{\sqrt{2}}\biggl(i\sin(\frac{\pi}{6}-\frac{\theta}{2}), \cos(\frac{\pi}{6}+\frac{\theta}{2}),-i\cos(\frac{\pi}{6}-\frac{\theta}{2}),-\sin(\frac{\pi}{6}+\frac{\theta}{2})\biggr)^T \\
    |\phi_2\rangle&=\frac{e^{-i\theta/2}}{\sqrt{2}}\biggl(-i\sin(\frac{\pi}{6}-\frac{\theta}{2}), \cos(\frac{\pi}{6}+\frac{\theta}{2}),i\cos(\frac{\pi}{6}-\frac{\theta}{2}),-\sin(\frac{\pi}{6}+\frac{\theta}{2})\biggr)^T.
\end{align}
We have chosen this basis because it yields particularly simple projections of the symmetry operations:
\begin{align}
    \langle\phi_i(\theta) | TR |\phi_j(\theta)\rangle &= s_x \\
    \langle\phi_i(\theta+\pi/3) | C_6 |\phi_j(\theta)\rangle &= \exp(i\pi s_z/3) \\
    \langle\phi_i(-\theta) | m_x |\phi_j(\theta)\rangle &= -s_x,
\end{align}
where we have introduced Pauli matrices $s_i$ acting in the space of $|\phi_i\rangle$. Using this basis, we can project the angular dispersion into the space of low-lying edge states to find the effective Hamiltonian
\begin{equation}
    \frac{1}{R}\langle\phi_i|i\sigma_y\tau_2(\theta)\partial_\theta|\phi_j\rangle=\frac{1}{R}(is_z\partial_\theta-\frac{1}{2}s_0),\label{eq:edgeham}
\end{equation}
which is the Hamiltonian for a pair of counter-propagating edge excitations. The term proportional to the identity accounts for the fact that our topological particle geometry has a constant-curvature edge\cite{wieder2020strong}; we will neglect it in the following as it does not contribute to our topological analysis.

At first glance, Eq.~\eqref{eq:edgeham} resembles the edge theory for the helical states of a two-dimensional topological insulator. In fact, the low-energy critical point Eq.~\eqref{eq:criticalham} coincides with the critical theory of a 2D TI. This observation led Wu and Hu to predict that topological particles such as ours should have a $\mathbb{Z}_2$ invariant with gapless counterpropagating edge states~\cite{Wu2015}. However, there is a fundamental distinction between our model and a two-dimensional TI due to the symmetries we require. To analyze the edge of our topological particle system, we should include higher-order terms in the bulk that preserve the 6mm point group symmetry, and ask what effect they have on the edge dispersion. Here, we will focus only on terms that cannot close a bulk gap, and that simultaneously gap the edge theory (\ref{eq:edgeham}). This means we look for potentials $V(\theta)$ that anticommute with both the bulk mass $m\Gamma_z$ and the edge kinetic term $\sigma_y\tau_2(\theta)$. However, we also require that $V(\theta)$ commute with $\Gamma_z\sigma_y\tau_1(\theta)$, in order that $\langle \phi_i | V(\theta) | \phi_j\rangle\neq 0$. We find that this restricts the form of $V(\theta)$ to 
\begin{equation}
    V(\theta)=m_4(\theta)\Gamma_4 + m_5(\theta)\Gamma_5,
\end{equation}
where we have introduced $\Gamma_5=i\Gamma_x\Gamma_y\Gamma_z\Gamma_4$. Crucially, both $\Gamma_4$ and $\Gamma_5$ anticommute with the sewing matrices for $C_6$ and $m_x$, we find:
\begin{subequations}
\begin{align}
    \langle \mathbf{w}_i | C_6 | \mathbf{w}_j\rangle &= \frac{1}{4}\left(\tau_0(\sigma_x-3i\sigma_y) + \sqrt{3}\tau_y(\sigma_y-i\sigma_x)\right),\\
     \langle \mathbf{w}_i | m_x | \mathbf{w}_j\rangle &= -\frac{1}{2}(\sigma_0\tau_z' - \sqrt{3}\sigma_z\tau_x').
\end{align}
\end{subequations}
Accounting for the action of the symmetries on the angular coordinate $\theta$, we can thus write a Fourier expansion
\begin{equation}
V(\theta)=\sum_n m_{4n}\sin((3+6n)\theta)\Gamma_4 + m_{5n}\cos((3+6n)\theta),
\end{equation}
where $n$ indexes the different Fourier harmonics. Projecting these onto the edge, we find that the edge Hamiltonian becomes
\begin{align}
    \nonumber H_\mathrm{edge} &= \frac{1}{R}(is_z\partial_\theta - 1/2s_0) + \sum_n \left[m_{4n}\sin((3+6n)\theta)\begin{pmatrix} 0 & ie^{-i\theta} \\ -ie^{i\theta} & 0 \end{pmatrix} + m_{5n}\cos((3+6n)\theta)\begin{pmatrix} 0 & e^{-i\theta} \\ e^{i\theta} & 0 \end{pmatrix}\right].
\end{align}
Let us focus on the case when only the $n=0$ masses are nonzero. To analyze this, we will without loss of generality take $m_{40}\neq 0$, $m_{50}=0$ to start, and then we will perturbatively reintroduce $m_{50}$:  the mass term $m_{40}\sin(3\theta)$ vanishes at the special values
\begin{equation}
    \theta_m = \frac{m\pi}{3}.
\end{equation}
Near each zero we have corner states which satisfy
\begin{equation}
    \frac{1}{R}\left[\partial_\theta +  3im_{40}(-1)^ms_z\theta\begin{pmatrix} 0 & ie^{-i\theta_m} \\ -ie^{i\theta_m} & 0 \end{pmatrix}\right]|\Theta_m\rangle=0,
\end{equation}
and so repeating our Jackiw-Rebbi analysis we find a zero-energy corner state satisfying
\begin{align}
    im_{40}(-1)^ms_z\theta\left(\begin{array}{cc} 0 & ie^{-i\theta_m}\nonumber \\ -ie^{i\theta_m} & 0 \end{array}\right)|\Theta_m\rangle = (-1)^{m+1}|\Theta_m\rangle,
\end{align}
yielding a total of six zero-energy corner states. We thus see that symmetry-allowed mass terms gap the counterpropagating edge states of Ref.~[\onlinecite{Wu2015}], yielding corner states consistent with our MPB and coupled dipole simulations. 

To complete the analysis, we next perturbatively restore $m_{50}$. Projecting into the space of corner modes for each $m$, we find
\begin{equation}
    m_{50}\cos(3\theta_m)\langle\Theta_m | \left(\begin{array}{cc} 0 & e^{-i\theta} \\ e^{i\theta} & 0 \end{array}\right) | \Theta_m\rangle = +m_{50}.
\end{equation}
This means that although $m_{50}$ breaks chiral symmetry and shifts the corner modes away from zero energy, it does not break the degeneracy of the corner modes. This leads to the so-called ``filling anomaly'': when both $m_{4}$ and $m_5$ are nonzero, the difference between the number of states in the positive and negative energy subspaces of the model is six. 

\begin{figure*}[!t]
\subfloat[]{\includegraphics[width=0.4\textwidth]{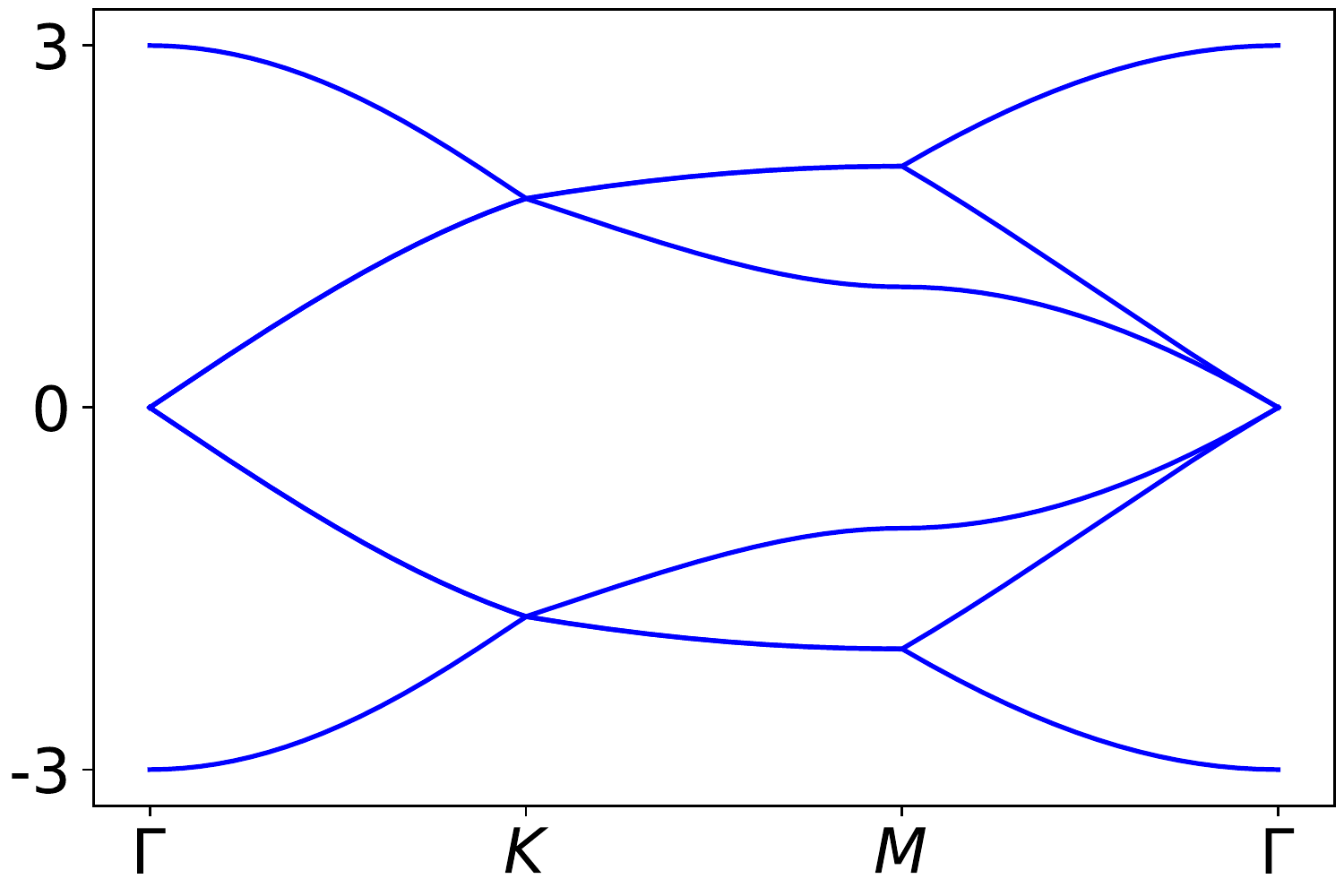}
}
\subfloat[]{\includegraphics[width=0.4\textwidth]{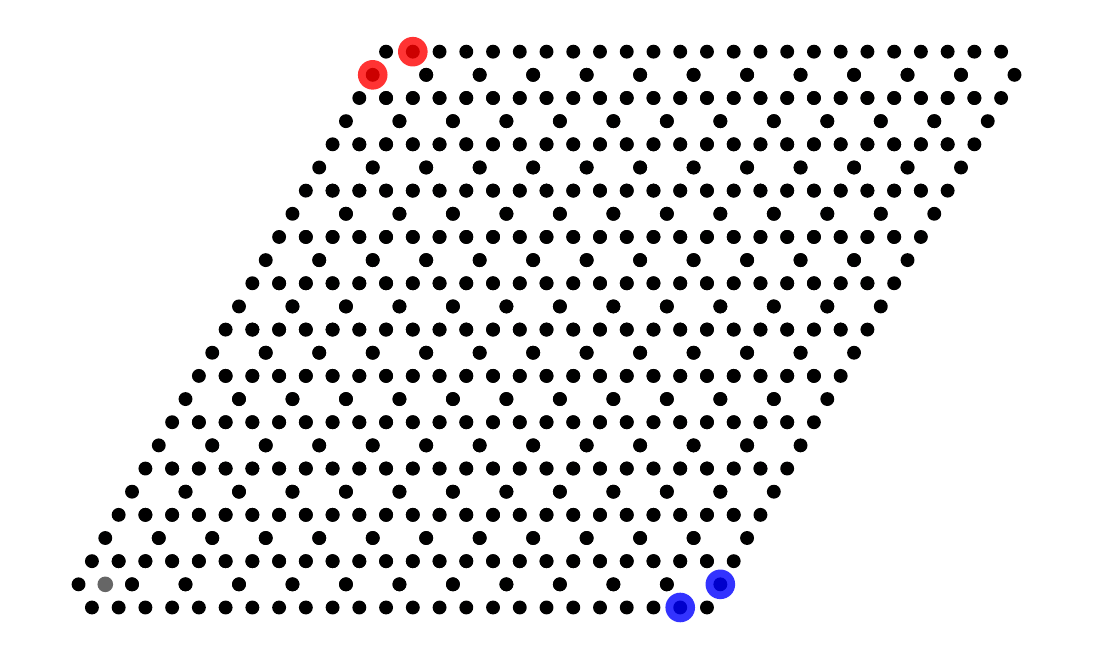}
}
\caption{(a) Band structure for the nanophotonic tight-binding model at the transition point between trivial and OAL phases. (b) Corner states for a $C_2$ symmetric topological particle in the OAL phase. The blue (red) circles represent the probability densities for the first (second) corner state.}\label{fig:tbcalc}
\end{figure*}

Note that we could have performed our same analysis with $m_5$ initially nonzero instead, which would result in corner modes localized at $\theta'_m = (2m+1)\pi/3$ (the other conjugacy class of mirror lines in the point group). Additionally, we could have considered higher Fourier harmonics in the mass term, which would yield additional sets of $12$ corner modes at generic points along the boundary, which gap non-anomalously. Finally, our analysis holds as well for a $C_2$-symmetric topological particle, in which case we can add mass terms of the form $\Gamma_5\cos 2\theta$ and $\Gamma_4\sin 2\theta$, which gap all but one pair of corner modes, yielding a filling anomaly of 2. We can see an example of this in the topological particle pictures in Fig.~\ref{fig:tbcalc}.

To conclude, let us comment on the applicability of our tight-binding calculation to the nanophotonic calculation. Because the full interaction matrix contains power-law decaying terms in position space, we cannot guarantee a priori that the Bloch Hamiltonian will permit a series expansion near the $\Gamma$ point in the Brillouin zone. However, for our model we find that the cusp singularities arising in the band structure due to the long-range hopping appear only in the highest positive and lowest negative energy bands in the band structure (one of which maps to the cusp singularity at $\omega=0$ in the full photonic model). Crucially, however, we have seen that it is only the bands close to the mid-gap band inversion that contribute to the formation of corner states in this model. Thus, we expect that our analysis here is robust to the inclusion of long range hoppings. It is an interesting open problem for future work to consistently incorporate band structure singularities due to long-range hoppings into the general theory of topological photonic systems.

\section{Exponential cutoff}

\begin{figure}[h]
    \centering
    \includegraphics[width=0.5\linewidth]{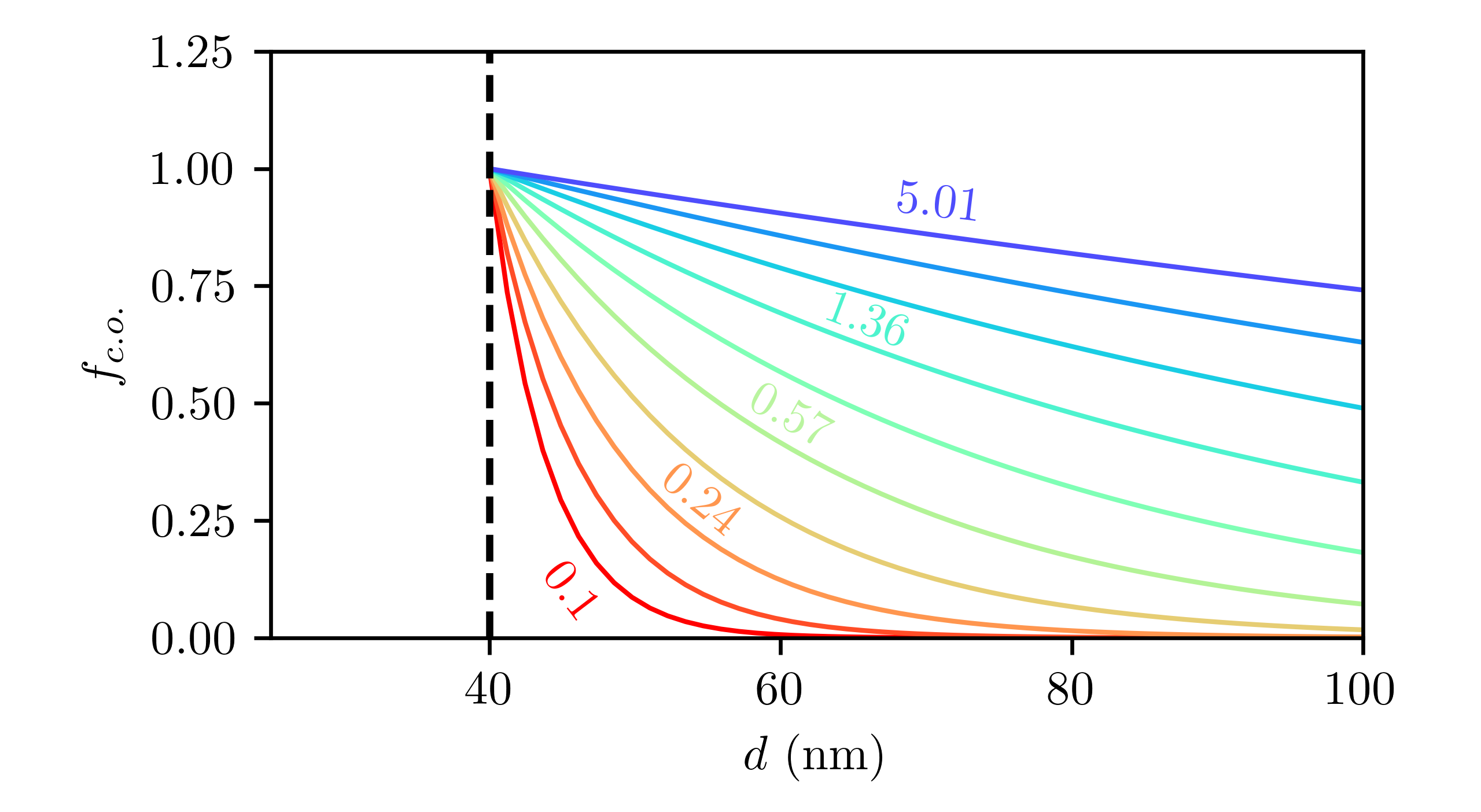}
    \caption{Example exponential cut off function, $f_{c.o.}$, for a nearest neighbour distance $d^0 = 40$~nm, for varying $\gamma$, from $0.1$ ($\log\gamma = -1$) (red) to $5.01$ ($\log\gamma = 0.7$) (blue). A cut off factor with $\gamma = 0.1$ for interactions $\propto -1/d^3$ is approximately nearest neighbour.}
    \label{fig:exp_cutoff}
\end{figure}

\section{Effect of disorder}

\begin{table*}[ht]
    \centering
    \begin{tabular}{|c|c|c|c|}
    \hline
         & $C_6$ breaking defects & Effect on topological corner states & Emergence of new states \\ \hline \hline 
         \multirow{5}{4em}{Corner defects} 
        & Remove 1 particle, “trivial” sublattice & 6 degenerate corner modes unaffected & one new localized state\\  \cline{2-4}
        & Remove 1 particle, “topological” sublattice & 5 degenerate corner modes w/o C6 symmetry & no \\  \cline{2-4}
        & Remove trimer at corner & 5 degenerate corner modes w/o C6 symmetry & no\\   \cline{2-4}
        & Expanded cell at corner & 5 degenerate corner modes w/o C6 symmetry &  yes, on both sublattices \\ \cline{2-4}
        & Contracted cell at corner & 5 degenerate corner modes w/o C6 symmetry & yes, on both sublattices  \\  \hline\hline
        \multirow{4}{4em}{Edge defects} 
        & One expanded cell at edge,   & 4+2 degenerate corner modes,  & yes, on both sublattices \\ 
        & preserving 1 mirror symmetry & w/ mirror symmetry &   \\ \cline{2-4}
        & One expanded cell at edge, & 4 degenerate corner modes w/ mirror symmetry,  & yes, on both sublattices \\ 
        &  breaking all mirror symmetries & + 2 non-degenerate w/o mirror symmetry & \\ \hline\hline
        \multirow{2}{4em}{Bulk defects} 
        & Random position disorder & 6 non-degenerate state,  & no \\ 
        &  & on “topological” sublattice & \\  \hline 
    \end{tabular}
    \caption{Effect of different $C_6$ breaking defects on the topological corner states for 
    particles with long range interactions. For a nearest neighbour model the corner states survive all perturbations except for the second corner defect type.  }
    \label{tab:disorder_summary}
\end{table*}

Table~\ref{tab:disorder_summary} summarizes the effect of different kinds of $C_6$ symmetry breaking defects on the topological corner modes of Types A and B particles: defects at corners, edges and random bulk disorder are considered. From the main text, the 6 degenerate corner modes survive when one particle belonging to the sublattice immediately at the corner is removed, even if $C_6$ symmetry is broken. All the other defects have an effect to some extent as shown in the table. In contrast, in a nearest neighbour model the topological corner modes are robust against all the perturbations considered in the table (except if one of the particles at the corner where the mode resides is removed). The effect on the topological robustness of the corner modes then emerges both from the spatial symmetries and the range of the interactions. At the critical point $s_*$, the Hamiltonian $H(\mathbf{k},s)$ is gapless with a fourfold Dirac degeneracy at the $\Gamma$ ($\mathbf{k}=0$) point.
These results hold true regardless of the edge termination, provided the particle has the same lattice symmetries. It should be noted that corner modes in 
particles with complete unit cells at the interface are more strongly affected by edge and bulk disorder, compared to the broken unit cell interface termination presented in the main text.
This is due to the longer localization length of these modes.

\clearpage
\bibliography{main.bib}